 \documentclass[authoryear,final,5p,times,twocolumn]{elsarticle}

\usepackage{graphicx}

\usepackage{natbib}

\def\Frac#1#2{{{\displaystyle\strut#1}\over{\displaystyle\strut#2}}}

\newcommand{\crm}{\cr\noalign{\medskip}}
\newcommand{\vv}[1]{\vec{#1}}

\newcommand{\be}{\begin{equation}}
\newcommand{\ee}{\end{equation}}
\newcommand{\ei}{\mathrm{e}}

\newcommand{\ii}{\mathrm{i}}
\newcommand{\jj}{\mathrm{j}}
\newcommand{\kk}{\mathrm{k}}

\newcommand{\KK}{\mathrm{K}}

\newcommand{\cV}{{\cal V}}

\newcommand{\iniper}{2}

\newcommand{\llabel}[1]{\label{#1}}

\newcommand{\figpath}{}
\newcommand{\vangle}{270}
\newcommand{\wh}{width}

\journal{Icarus}

\begin{document}

\begin{frontmatter}

\title{Mercury's capture into the 3/2 spin-orbit resonance \\
including the effect of core-mantle friction}

\author[lab1,lab2]{Alexandre C. M. Correia\corref{cor1}}
\ead{correia@ua.pt}

\cortext[cor1]{Corresponding author, }

\address[lab1]{Departamento de F\'\i sica, Universidade de Aveiro,
Campus de Santiago, 3810-193 Aveiro, Portugal}

\author[lab2]{Jacques Laskar}

\address[lab2]{Astronomie et Syst\`emes Dynamiques, IMCCE-CNRS UMR8028, Observatoire
de Paris, UPMC, 77 Av. Denfert-Rochereau, 75014 Paris, France}



\begin{abstract}

The rotation of Mercury is presently captured in a 3/2 spin-orbit resonance with the
orbital mean motion.
The capture mechanism is well understood as the result of tidal interactions
with the Sun combined with planetary perturbations \citep{Goldreich_Peale_1966, Correia_Laskar_2004}.
However, it is now almost certain that Mercury has a liquid core \citep{Margot_etal_2007} 
which should induce a contribution of
viscous friction at the core-mantle boundary to the spin evolution.
According to \citet{Peale_Boss_1977} 
this last effect greatly increases the
chances of capture in all spin-orbit resonances, being 100\% for the 2/1 resonance,
and thus preventing the planet from evolving to the presently observed configuration.
Here we show that for a given resonance, as the chaotic evolution of Mercury's orbit can drive its
eccentricity to very low values during the planet's history, any
previous capture can be destabilized whenever the eccentricity becomes lower than a critical value. 
In our numerical integrations of 1000 orbits of Mercury over 4\,Gyr,
the spin ends 99.8\% of the time captured in a spin-orbit resonance, 
in particular in one of the following three configurations: 5/2
(22\%), 2/1 (32\%) and 3/2 (26\%).
Although the present 3/2 spin-orbit resonance is not the most probable outcome, 
we also show that the capture probability in this resonance can be increased up
to 55\% or 73\%, if the eccentricity of Mercury in the past has descended
below the critical values 0.025 or 0.005, respectively.

\end{abstract}

\begin{keyword}
Mercury \sep spin dynamics \sep tides \sep 
core-mantle friction \sep resonance
\end{keyword}

\end{frontmatter}

\section{Introduction}
\label{intro}

Mercury's present rotation is locked in a 3/2 spin-orbit resonance, 
with its spin axis nearly perpendicular to the orbital plane
\citep{Pettengill_Dyce_1965}.
The stability of this rotation results from the solar torque on Mercury's
quadrupolar moment of inertia, combined with an eccentric orbit:
the axis of minimum moment of inertia is always aligned with the direction to
the Sun when Mercury is at the perihelion of its orbit
\citep{Colombo_1965,Goldreich_Peale_1966,Counselman_Shapiro_1970}.  
The initial rotation of Mercury was presumably faster than today, but tidal
dissipation along with core-mantle friction brought the planet rotation to the 
present configuration, where capture can occur.
However, the exact mechanism on how this state initially arose is not completely
understood. 

In their seminal work, \citet{Goldreich_Peale_1966} have shown that 
since the tidal strength depends on the planet's rotation rate, it creates an
asymmetry in the tidal potential that allows capture into 
spin-orbit resonances. They also computed the capture probability 
into these resonances for a single crossing, 
and found that for the present eccentricity value of Mercury ($ e = 0.206 $),
and  unless one uses an unrealistic tidal model with constant torques (which
cannot account for the observed damping of the planet's libration), the
probability of capture into the present 3/2 spin-orbit resonance is on the low
side, at most about 7\%, which  remained somewhat unsatisfactory. 

\citet{Goldreich_Peale_1967} nevertheless pointed out that the
probability of capture could be greatly enhanced if a planet has a molten core.
In 1974, the discovery of an intrinsic magnetic field by the Mariner~10
spacecraft \citep{Ness_etal_1974}, seemed to imply the existence of a
conducting liquid core and consequently an increment
in the capture probability in the 3/2 resonance.
However, according to \citet{Goldreich_Peale_1967}, 
this also increases the capture probability  in all the previous resonances.
\citet{Peale_Boss_1977} indeed remarked  that only very specific values of the
core viscosity allow to avoid the 2/1 resonance and permit the capture in the
3/2 configuration. 

More recently, \citet{Correia_Laskar_2004} (hereafter denoted by Paper~I) have
shown that as
the orbital eccentricity of Mercury is varying chaotically, 
from near zero to more than 0.45, the capture probability is substantially increased.
Indeed, when the large eccentricity variation is factored into the
capture, the rotation rate of the planet can be accelerated, and 
the 3/2 resonance could have been crossed many times in the past. 
Performing a statistical study of the past evolutions of Mercury's orbit,
over 1000 cases, it was demonstrated that
capture into the 3/2 spin-orbit resonant
state is in fact, and without the need 
of specific core-mantle effect, the most probable final outcome of the planet's evolution,
occurring about 55.4\% of the time.
In contrast, because the eccentricity can decrease to near zero, 
all resonances except the 1/1 become unstable, allowing
the planet to escape from resonance. 
This mechanism suggests that in presence of core-mantle friction
the planet can escape to a previous 
capture in the 2/1 or higher order spin-orbit resonances. 

The present paper continues the work started in Paper~I.
In addition to the effects of tides and planetary perturbations, we will
consider here also the core-mantle friction effect as described by
\citet{Goldreich_Peale_1967}, since the presence of a
liquid core inside Mercury is now confirmed by radar observations \citep{Margot_etal_2007}. 
In the next section we give the averaged equations of motion in a suitable
form for simulations of the long-term variations of Mercury's spin, including
the resonant motion, viscous tidal effects, core-mantle coupling and
planetary perturbations.
In section~3 we discuss the consequences of each effect into the spin evolution
and evaluate the capture probabilities in resonance.
Finally, in last section we perform some numerical simulations to illustrate
the different effects described in section~3.

\section{Equations of motion}

\llabel{eqomt}

We will adopt here a model for Mercury which is an extension of the model from
\citet{Poincare_1910} of a perfect incompressible and homogeneous liquid core
with moments of inertia $ A_c = B_c < C_c $ inside an homogeneous
rigid body with moments of inertia $ A_m \le B_m < C_m $, supported by the
reference frame $ ( \vv{\ii}, \vv{\jj}, \vv{\kk} ) $, fixed with respect to the
planet's figure. 
Tidal dissipation and core-mantle friction drive the obliquity close to
zero \citep{Yoder_1997,Correia_etal_2003}. 
Since we are only interested here in the study of the final stages of
evolution (where capture in spin-orbit resonance may occur), we will neglect the effect of
the small obliquity variations on the equations of motion.
Moreover, since for a long-term study we are not interested in diurnal nutations, we can average
over fast rotation angles and merge the axis of principal
inertia and the axis of rotation \citep{Boue_Laskar_2006}. 
Therefore, the mantle rotation rate is simply given by $ \omega = \dot \theta + \dot
\varphi$, where $ \theta $ is the rotation angle and $ \varphi $ the precession
angle.

\subsection{Precession torque}

\llabel{070730a}

The gravitational potential $ \cV $ generated at a generic point of the space $
\vv{r} $ is given by \citep[e.g.][]{Tisserand_1891,Smart_1953}: 
\begin{eqnarray}
\cV (\vv{r}) = - \Frac{G m}{r} & + & \Frac{G (B - A)}{r^3} P_2 ( \vv{u}_r
\cdot \vv{\jj} ) \crm & + &  \Frac{G (C - A)}{r^3} P_2 ( \vv{u}_r \cdot
\vv{\kk} ) \llabel{070730b} \ ,
\end{eqnarray}
where  terms in $ (R/r)^3 $ were neglected.
$ G $ is the gravitational constant, $ m $ the mass of Mercury, $ \vv{u}_r =
\vv{r} / r $ and $ P_2 (x) = (3 x^2 -1)/2 $ is the Legendre 
polynomial of degree two. 
When interacting with the Sun's mass, $ m_\odot $, the spin of Mercury will
undergo important changes.
The middle term in the above potential will be responsible for a libration in
the spin, while the last term causes the spin axis $ \vv{\kk} $ to precess
around $ \vv{\KK} $, the normal to the orbit.
Since we are only interested in the study of the long term motion, we 
will average the potential over the rotation angle $\theta$ and the
mean anomaly $ M $, after expanding the true anomaly $ v $ in series of the
eccentricity $e$ and mean anomaly. 
However, 
when the rotation rate $ \omega $ and the mean motion $ n = \dot M $
are close to resonance ($ \omega \simeq p n $, for a semi-integer\footnote{We have retained the use
of semi-integers for better  comparison with previous results.} value $p$),
we must retain the terms with argument $ (2 \theta - 2 p M) $ in the expansion
\be 
\frac{\cos ( 2 \theta - 2 v )}{r_{}^3} = \frac{1}{a^3}
\sum^{+\infty}_{p=-\infty} H ( p , e ) \cos ( 2 \theta - 2 p M) \ ,
\llabel{061120gb} 
\ee  
where  $ a $ is the semi-major axis of the planet's orbit and 
the function $ H (p, e) $ is a power series in $ e $ (Tab.~\ref{TAB1}).
\begin{table*}
\caption{Coefficients of $ H ( p , e) $ to  $ e^7 $.
The exact expression of these coefficients is given by $  H ( p , e) =
\frac{1}{\pi} \int_0^\pi \left( \frac{a}{r} \right)^3 \exp(\ii \, 2 \nu)
\exp(\ii \, 2 p M) \, d M $. \llabel{TAB1} } 
\begin{center}
\begin{tabular}{| c |cccccccc|} \hline 
$ \quad p \quad $ & \multicolumn{8}{c|}{$ \phantom{\Frac{1}{1}} H (p,e) \phantom{\Frac{1}{1}} $} \\ \hline 
1/1 & 1 & $-$ & $\Frac{5}{2} e^2$ & $+$ & $\Frac{13}{16} e^4$ & $-$ & $\Frac{35}{288} e^6$ &  \\ \hline
3/2 & & $\Frac{7}{2} e$ & $-$ & $\Frac{123}{16} e^3$ & $+$ & $\Frac{489}{128} e^5$ & $-$ & $\Frac{1763}{2048} e^7$ \\ \hline
2/1 & & & $\Frac{17}{2} e^2$ & $-$ & $\Frac{115}{6} e^4$ & $+$ & $\Frac{601}{48} e^6$ &  \\ \hline
5/2 & & & & $\Frac{845}{48} e^3$ & $-$ & $\Frac{32525}{768} e^5$ & $+$ & $\Frac{208225}{6144} e^7$  \\ \hline
3/1 & & & & & $\Frac{533}{16} e^4$ & $-$ & $\Frac{13827}{160} e^6$ &  \\ \hline
7/2 & & & & & & $\Frac{228347}{3840} e^5$ & $-$ & $\Frac{3071075}{18432} e^7$  \\ \hline
4/1 & & & & & & & $\Frac{73369}{720} e^6$ &  \\ \hline
9/2 & & & & & & & & $\Frac{12144273}{71680} e^7$  \\ \hline
\end{tabular} \end{center}
\end{table*}
The exact averaged contributions to the spin are given in a suitable form for
our study by expression (15) in \citet{Correia_2006}\footnote{There is a misprint
in the sign of $ \phi $ in \citet{Correia_2006}.}. 
For zero obliquity we have:
\be
\Frac{d \omega}{d t} = - \frac{\beta}{c_m} \, H(p,e) \sin 2 (\theta + \varphi - p M -
\varpi) \ , \llabel{070730c} 
\ee
where $ \varpi $ is the longitude of the
perihelion, $ c_m = C_m / C = 0.55 $ \citep{Margot_etal_2007}, and
\be 
\beta = \Frac{3 G m_\odot}{2 a^3} \Frac{B - A}{C} \simeq \Frac{3}{2}
n^2 \Frac{B - A}{C} \ . \llabel{030123g}
\ee

The Mariner~10 flyby of Mercury provided information on the internal
structure of the planet, though subject to some uncertainty.
For the gravity field it has been measured $
J_2 = (6.0 \pm 2.0) \times 10^{-5} $ and $ C_{22} = (1.0 \pm 0.5) \times
10^{-5} $ \citep{Anderson_etal_1987}. 
Modeling the interior structure of Mercury, it has been estimated for the
structure constant $ \xi = C / ( m R^2 ) \simeq 0.3359 $ \citep{Spohn_etal_2001}. 
Thus, for the moments of inertia we compute $ (B-A)/C = 4 \, C_{22} / \xi \simeq
1.2 \times 10^{-4} $.

\subsection{Tidal torques} 

\llabel{070731a}

Tidal effects arise from differential and inelastic deformations of Mercury
due to the gravitational effect of the Sun. 
Their contributions to the spin variations are based on a very
general formulation of the tidal potential, initiated by George H. \citet{Darwin_1880}. 
The distortion of the planet gives rise to a tidal potential,
\be 
\cV^g (\vv{r},\vv{r}') = - k_2 \frac{G m_{\odot}}{R} \left( \frac{R}{r_{}}
\right)^3 \left( \frac{R}{r'} \right)^3 P_2 \left( \vv{u}_r \cdot 
\vv{u}_{r'} \right) \ . \llabel{070731b} 
\ee 
where $ R $ is the average radius of the planet and
$ \vv{r'} $ the radial distance from the planet's center to the Sun. 
In general, imperfect elasticity will cause the phase angle of $ \cV^g $ to lag
behind the perturbation \citep{Kaula_1964}, because there is is a time delay $
\Delta t $ between the perturbation of the Sun and the maximal deformation of
Mercury.
During that time, the planet rotates by an angle $ \omega \Delta t $, while
the Sun also changes its position. 
Assuming a constant time delay allows us to linearize the tidal potential and
simplify the tidal equations \citep{Mignard_1979,Mignard_1980,Hut_1981}.
For zero obliquity, the averaged contributions to the spin are then
given by:
\be
\Frac{d \omega}{d t} = - \frac{K}{c_m} \left[ \Omega (e) \Frac{\omega}{n} - N
(e) \right] \ , \llabel{070731c}
\ee
where
\be
\Omega (e) = \Frac{1 + 3 e^2 + 3 e^4 / 8}{(1 - e^2)^{9/2}} \ , \llabel{030218a}
\ee
\be
N (e) = \Frac{1 + 15 e^2 / 2 + 45 e^4 / 8+ 5 e^6 / 16}{(1 - e^2)^{6}} \ ,
\llabel{030218b}
\ee
\be
K = n^2 \, \frac{3 \, k_2}{\xi \, Q} \left(\frac{m_\odot}{m}\right) 
\left(\frac{R}{a}\right)^3 \ , \llabel{eq3}
\ee
and $ Q^{-1} = n \Delta t $.
This tidal model is particularly adapted to describe the planets behavior in
slow rotating regimes ($ \omega \sim n $), which is the case of Mercury during
the spin-orbit resonance crossing.
In the present work we will adopt $ k_2 = 0.4 $ and $ Q = 50 $, which yields $ K
= 2.2 \times 10^{-5} \, \mathrm{yr}^{-2} $.
This choice is somewhat arbitrary, but based on the parameter 
values of the other terrestrial
planets \citep{Goldreich_Soter_1966}.

\subsection{Core-mantle friction effect}

\llabel{sectioCoreMantle}

The Mariner~10 flyby of Mercury revealed the presence of an intrinsic magnetic
field, which is most likely due to motions in a conducting fluid core
\citep[for a review see][]{Ness_1978}.
Subsequent observations made with Earth-based radar provided strong evidence that
the mantle of Mercury is decoupled from a core that is at least
partially molten \citep{Margot_etal_2007}.

If there is slippage between the liquid core and the mantle, a second source of
dissipation of rotational energy results from friction occurring at the
core-mantle boundary. 
Indeed, because of their different shapes and densities, the core and the mantle
do not have the same dynamical ellipticity and the two parts tend to precess
at different rates \citep{Poincare_1910}.
This tendency is more or less counteracted by different interactions produced at
their interface: the torque of non-radial inertial pressure forces of
the mantle over the core provoked by the non-spherical shape of the interface;
the torque of the viscous friction (or turbulent) between the core and the
mantle; and the torque of the electromagnetic friction, caused by the interaction
between electrical currents of the core and the bottom of the magnetized mantle.
The two types of friction torques (viscous and electromagnetic) depend
on the differential rotation between the core and the mantle and can be
expressed by a single effective friction torque, $ \Upsilon $
\citep{Mathews_Guo_2005,Deleplace_Cardin_2006}:
\be
\Upsilon = C_c \kappa \, \delta \ ; \quad \delta = \omega - \omega_c \ ,
\llabel{070731h}
\ee
where $ \kappa $ is an effective coupling parameter, $ \omega_c $ the core's
rotation rate and $ c_c = C_c / C = 1 - c_m = 0.45 $ \citep{Margot_etal_2007}.
According to \citet{Mathews_Guo_2005} we may write $ \kappa \simeq 2.62
\sqrt{\nu \omega} / R_c $, where $ R_c $ is the core radius and $
\nu $ is the effective kinematic viscosity of the core.
Adopting $ R_c / R = 0.77 \pm 0.04 $ \citep{Spohn_etal_2001} and $ \nu = 10^{-6}
\mathrm{m}^2 \mathrm{s}^{-1} $, we compute for the 3/2 resonance: $
\kappa = 5 \times 10^{-5} \, \mathrm{yr}^{-1} $.
The uncertainty over $ \nu $ is very large, according to
\citet{Lumb_Aldridge_1991} it can cover about 13 orders
of magnitude. 
As in former studies on planetary evolution \citep{Yoder_1997,
Correia_Laskar_2001, Correia_Laskar_2003I}, we used the same value as the best
estimated so far for the Earth \citep{Gans_1972,Poirier_1988}.

Another important consequence of core-mantle friction is to tilt the equator of the
planet to the orbital plane, which results in a secular decrease of the
obliquity \citep{Rochester_1976,Correia_2006}, reinforcing the previous
assumption that the obliquity remains close to zero during the last
evolutionary stages.

Since the core and the mantle are decoupled, we need a differential equation for
each rotation rate, the coupling equation being given by the friction torque
\citep{Peale_2005,Correia_2006}:
\be
\Frac{d \omega}{d t} = - \frac{c_c \kappa}{c_m} \, \delta \ \quad \mathrm{and}
\quad \Frac{d \omega_c}{d t} = \kappa \, \delta \ . \llabel{070731i}
\ee

\subsection{Planetary perturbations}

\llabel{070802a}

When considering the perturbations of the other planets, the eccentricity of
Mercury is no longer constant, but undergoes strong chaotic variations in time
\citep{Laskar_1990,Laskar_1994,Laskar_2008}.
These variations can be modeled using the averaging of the equations for the motion of
the Solar System, that have been compared to
recent numerical integrations, with very good agreement \citep{Laskar_etal_2004M,
Laskar_etal_2004E}.
These equations are
obtained by averaging the equations of motion over the
fast angles that are the mean longitudes of the planets.
The averaging of the equation of motion is obtained by
expanding the perturbations of the Keplerian orbits in
Fourier series of the angles, where the coefficients themselves
are expanded in series of the eccentricities and inclinations.
This averaging process was conducted in a
very extensive way, up to second order with respect to
the masses, and through degree five in eccentricity and inclination,
leading to truncated secular equations of the
Solar System of the form
\be
\frac{d w}{d t} = \ii \left[ \Gamma w + \Phi_3 (w, \bar w) + \Phi_5 (w, \bar w)
\right] \ , \llabel{080404b}
\ee
where $ w$ is the column vector  $ (z_1,...,z_8,\zeta_1,...,\zeta_8) $, with
$ z_j = e_j \exp (\ii \varpi_j) $, $ \zeta_j = \sin (I_j/2) \exp (\ii \Omega_j)
$ and $ \varpi $ and $ \Omega $, respectively the longitude of the perihelion
and node. 
The $ 16 \times 16 $ matrix $\Gamma$ is the linear Lagrange-Laplace system,
while $ \Phi_3 (w, \bar w) $ and $ \Phi_5 (w, \bar w) $ gather terms of degree 3
and 5 respectively.

The system of equations thus obtained contains some
150000 terms, but  its main frequencies are now the precession frequencies
of the orbits of the planets. The full system can thus be
numerically integrated with a very large step-size of 250
years. Contributions due to the secular perturbation
of the Moon and general relativity are also included \citep[for more details and
references see][]{Laskar_1990,Laskar_1996,Laskar_etal_2004M}.

This secular system is then simplified and reduced to
about 50000 terms, after neglecting terms of very small
value \citep{Laskar_1994}.
Finally, a small correction of the terms appearing in $\Gamma$
(Eq.\ref{080404b}), after diagonalization, is performed 
in order to adjust the linear frequencies, in a similar
way as it was done by \citet{Laskar_1990}. 
The
secular solutions are very close to the direct numerical integration
La2004 \citep{Laskar_etal_2004M, Laskar_etal_2004E} over about 35\,Myr, 
the time over which the direct numerical solution
itself is valid because of the imperfections of the model. 
It is thus legitimate to investigate the diffusion of the orbital motion of
Mercury over long times using the secular equations.

Over several millions years, the eccentricity of Mercury presents 
significant variations (Fig.\ref{ecmerc}) that need to be taken into account 
in the resonance capture process (Paper~I).


\section{Spin evolution and capture probabilities}

\llabel{070803a}

The evolution of the spin for zero obliquity is completely described when
we put together the contributions from the different effects presented in
section~\ref{eqomt}:
\be
 \left\{ 
  \begin{array}{l l}
   \dot \omega = P_{} + T_{} - \Upsilon / C_m \ , \crm
   \dot \omega_c = \Upsilon / C_c \ , 
  \end{array} 
 \right.
\llabel{070803b}
\ee 
where $ P_{} $ is the precession torques (Eq.\ref{070730c}), $ T_{} $ 
the tidal torque (Eq.\ref{070731c}) and  $ \Upsilon $ 
the friction torque (Eq.\ref{070731h}).
In order to better study the capture probabilities in spin-orbit
resonances, we will adopt a  change of variables that is valid around each
resonance $ p $:
\be
\gamma = \theta + \varphi - p M - \varpi \ , \llabel{070803c1} 
\ee
and thus,
\be
\dot \gamma = \omega - p n - \dot \varpi \quad \mathrm{and} \quad \ddot \gamma =
\dot \omega - \ddot \varpi \ . \llabel{070803c2} 
\ee
In absence of planetary perturbations, for a Keplerian orbit,  $ \dot \varpi = 0 $
and thus $ \ddot \gamma = \dot \omega $.
Instead of the core rotation $ \omega_c $ we will also adopt the differential
rotation $ \delta = \omega - \omega_c $ as variable. 
The above system of equations (\ref{070803b}) becomes then:
\be
 \left\{ 
  \begin{array}{l l}
   \ddot \gamma = P_{}(\gamma) + T_{} (\dot \gamma) - c_c \kappa_m
   \delta - \ddot \varpi \ , \crm
   \dot \delta = P_{}(\gamma) + T_{} (\dot \gamma) - \kappa_m \delta \ ,
  \end{array} 
 \right.
\llabel{070803d}
\ee
where $ \kappa_m = \kappa / c_m $.
In general, we like to express $ \ddot \gamma $ as the sum of
the precession torque $ P_{} (\gamma) $ and a global dissipative torque $
D (\dot \gamma) $ such that
\be
\ddot \gamma = - \beta_m \sin 2 \gamma + D (\dot \gamma) \ ,
\llabel{070803e} 
\ee
where
\be
\beta_m = \frac{\beta}{c_m} H(p,e) \llabel{070825a} 
\ee
and
\be
D (\dot \gamma) = - D_0 \left( V + \mu \Frac{\dot \gamma}{n} \right) \ ,
\llabel{070803f} 
\ee
$ D_0 $, $ V $ and $ \mu $ being ``constant'' quantities.
Indeed, under this linearized form we are able to estimate the capture
probabilities using a simple expression derived by \citet{Goldreich_Peale_1966}:
\be 
P_\mathrm{cap} = 2 \left[ 1 + \frac{\pi}{2} \frac{n}{\Delta \omega}
\frac{V}{\mu} \right]^{-1} \ , \llabel{070803g} 
\ee
where $ \Delta \omega $ is the maximal amplitude of libration in resonance,
i.e., $ \Delta \omega = \sqrt{2 \beta_m} $.

\subsection{Effect of tides}

\llabel{070803z}

Let us consider first the simplified case where the spin of the planet is only
subject to the precession and tidal torques \citep{Goldreich_Peale_1966}.
Thus, we may use expression (\ref{070803e}) to express $ \ddot \gamma $, where 
$ D (\dot \gamma) $ is given by expression (\ref{070731c}):
\be
D (\dot \gamma) = T_{} = - \frac{K}{c_m} \Omega (e) \left[ p 
- E (e) + \Frac{\dot \gamma}{n} \right] \ ,
\llabel{070803h}  
\ee
with $ E(e) = N(e) / \Omega(e) $.
The limit solution of the rotation for a constant value of the eccentricity is
obtained when $ D (\dot \gamma) = 0 $, that is, for $ \dot \gamma / n = E(e) - p
\Leftrightarrow \omega / n = E(e) $.
If the planet encounters a spin-orbit resonance in the way to the equilibrium
position, capture may occur with a probability computed from expression (\ref{070803g}):
\be 
P_\mathrm{cap} = 2 \left[ 1 + \frac{\pi}{2} \frac{n}{\Delta \omega}
\left( p - E (e) \right) \right]^{-1} \ . \llabel{070803i} 
\ee
Using the present value of the eccentricity of Mercury and $ c_m = 1 $ we
compute for the $ p = 3/2 $ resonance a probability of capture of about 7\%,
which is unsatisfactory given that this is the presently observed resonant state.


\subsection{Effect of core-mantle friction}

\llabel{070803x}

We now add the effect of core-mantle friction to the effects already considered 
in the previous section.
In this case we must take into account the rotation of the core, and the
complete rotation rate evolution is given by system (\ref{070803d}) with $ \ddot
\varpi = 0 $.

The general solution for the differential rotation in system (\ref{070803d}) is
given by:
\be
\delta (t) = \delta_0 \, \ei^{-\kappa_m t} + \ei^{-\kappa_m t} \int_{t_0}^t
\left[ P_{} (\gamma) + T_{} (\dot \gamma) \right] \ei^{\kappa_m t'} d t'
 \ , \llabel{061116a}    
\ee
where $ \delta_0 $ is the initial value of $ \delta (t) $ for $ t = t_0 $.
In section~\ref{sectioCoreMantle} we estimate for the present rotation of
Mercury $ 1 / \kappa_m \simeq 10^4 $\,yr, which can be seen as the 
time scale needed to damp the initial differential rotation $ \delta_0 $ to zero.
Thus, for $ t \gg 1/ \kappa_m $ the evolution of the differential rotation is
given only by the last term is expression (\ref{061116a}).
It also means that for time intervals $ \Delta t \ll 1 / \kappa_m $, we can write
for $ t < t_0 + \Delta t $:
\be
\delta (t) \simeq \delta_0 + \int_{t_0}^t \ddot \gamma \ d t' \ ,
\llabel{070804d1} 
\ee
that is,
\be 
\delta (t) \simeq \delta_0 + \dot \gamma - \dot \gamma_0 = \dot \gamma -
\dot \gamma_{c_0} \ , \llabel{070804d2}
\ee
where $ \dot \gamma_{c_0} = \dot \gamma_0 - \delta_0 = \omega_{c_0} - p n $, and
$ \omega_{c_0} $ is the value of the core rotation for $ t = t_0 $. 
This approximation is very useful because we can express $ \ddot \gamma $ by
means of expression (\ref{070803e}) with
\begin{eqnarray}
D (\dot \gamma) & = & - \frac{K}{c_m} \Omega (e) \left[  p 
 - E (e) + \Frac{\dot \gamma}{n} \right] - \Frac{c_c
\kappa}{c_m} \left( \dot \gamma - \dot \gamma_{c_0} \right) \crm
& = & - \frac{K}{c_m} \Omega (e) \left[ p - E (e) -
\chi \Frac{\dot \gamma_{c_0}}{n} + ( 1 + \chi) \Frac{\dot \gamma}{n} 
\right] \ , \llabel{070804e}  
\end{eqnarray}
where $ \chi = c_c \kappa n / [ K \Omega(e) ] $.
It allows us easily to compute the capture probabilities using expression
(\ref{070803g}) provided that we are able to estimate correctly $ \dot
\gamma_{c_0} $ just before the planet crosses the resonance:
\be 
P_\mathrm{cap} = 2 \left[ 1 + \frac{\pi}{2} \frac{n}{\Delta \omega}
\Frac{p - E (e) - \chi \Frac{\dot \gamma_{c_0}}{n}}{1+\chi}
\right]^{-1} \ . \llabel{070806a}
\ee

According to expression (\ref{070731i}) we have $ \dot \omega_c = \kappa \,
\delta $. 
Since $ \kappa \ll \Delta \omega $, the core is unable to follow the
periodic variations in the mantle's rotation rate.
Thus, only the secular variations on the mantle will be followed by the core and
we may write: $ \omega_c = \overline{\omega} - \overline{\delta} $, where $
\overline{\omega} $ is
the averaged rotation of the mantle and
\be
\overline{\delta} \simeq \ei^{-\kappa_m t} \int_{t_0}^t T_{} \ \ei^{\kappa_m t'} d
t' \simeq \Frac{T_{}}{\kappa_m} \ . \llabel{070806d}   
\ee
Just before crossing the $ p^\mathrm{th} $ spin-orbit resonance we have $
\overline{\omega_0} = p n + 2 \Delta \omega / \pi $, and thus
\begin{eqnarray}
\dot \gamma_{c_0} & = & \omega_{c_0} - p n  = \overline{\omega_0} -
\overline{\delta} - pn  \crm & = & \frac{2}{\pi} \Delta \omega -
\Frac{T_{}}{\kappa_m} \simeq \frac{2}{\pi} \Delta \omega + \frac{K \Omega
(e)}{\kappa} \left[ p  - E (e) \right]\ . \llabel{070806e}  
\end{eqnarray}
When using the present eccentricity of Mercury and the values of $ \kappa $ and
$ K $ estimated in previous sections, we compute $ \chi = 19.5 $.
Substituting this into expression (\ref{070806a}) we obtain a probability of capture of
100\% in the 3/2 spin-orbit resonance.
However, as noticed by \citet{Peale_Boss_1977}, if we compute the probability for
the 2/1 resonance we also get 100\%. 
Thus, either the planet started its rotation below the 2/1 resonance, which is
unlikely, or there must be another mechanism to avoid capture in the 2/1
and higher order resonances.

\subsection{Effect of planetary perturbations}

\llabel{070803y}

The orbital eccentricity of Mercury undergoes important secular perturbations
from the other planets (Fig.\ref{ecmerc}) and its contribution needs to be taken
into account. 
The mean value of the eccentricity is $ \bar e = 0.198 $, slightly lower than the
present value $ e \simeq 0.206 $, but we also observe a wide range for
the eccentricity variations, from nearly zero to more than 0.45
\citep[Paper~I,][]{Laskar_2008}. 
Even if some of these episodes do not last for a long time, they will allow additional
capture into and escape from spin-orbit resonances.
Moreover, the capture probabilities are also modified for
different eccentricities: for the same resonance we can have zero or 100\% of
captures depending on the eccentricity value (Fig.\ref{probe}).
For all resonances, the capture probability is 100\% whenever the eccentricity is
close to the equilibrium value for the rotation rate, $ E(e) = N(e) / \Omega(e) $ 
(Eq.\ref{070803h}), but it tends to decrease as the eccentricity moves away
from this equilibrium value.
If the eccentricity is too high (or too low if the spin is increasing from lower
rotation rates) some resonances cannot be reached and the probability of
capture suddenly drops to zero (Fig.\ref{probe}).

For a non-constant eccentricity $e(t)$, the limit solution of the rotation rate
when $ D (\dot \gamma) = 0 $ (Eq.\ref{070803e})
is no longer $ \omega / n = E(e) $, but more generally:
\be
\omega (t) = \frac{1}{g(t)} \int_0^t  \frac{K}{c_m}
\left[ N(e(\tau)) - \frac{c_c \kappa}{K} \delta (\tau) \right] g(\tau) \, d
\tau \ , \llabel{080306a} 
\ee
where
\be
g(t) = \exp \left( \frac{K}{c_m n} \int_0^t \Omega(e(\tau)) \, d \tau \right) 
\ . \llabel{080306b}
\ee
The dissipation torques can thus drive the rotation rate several times across the
same spin-orbit resonance, increasing the chances of capture.

Another important consequence of a non-constant eccentricity is that all
resonances but the 1/1 may become unstable.
Indeed, the amplitude of the libration torque depends on the coefficient $ H(p,
e) $ (Eq.\ref{070730c}), which goes to zero with the
eccentricity, except for the 1/1 resonance (Tab.\ref{TAB1}).
Whenever the amplitude of the libration restoration torque becomes smaller than
the amplitude of the dissipation torque, equilibrium in the spin-orbit resonance
can no longer be sustained and the resonance is destabilized.
Critical eccentricities for each resonance are listed in Table~\ref{Tabb},
obtained when the torques become equivalent (Eqs.\ref{070730c},\ref{070731c}):
\be
\frac{ \Omega (e) p - N (e) }{H (p, e)} = \frac{\beta}{K} \ . \llabel{080319a}
\ee

\begin{table}
\caption{Critical eccentricity $e_c$ for the resonance $p$. If 
$e < e_c$, the resonance $p$ becomes unstable, and the solution may escape
the resonance (Paper~I).
The critical eccentricity $e_c$ is obtained when
$ \beta H(p,e) < K [ \Omega (e) p - N (e) ] $.\llabel{Tabb}} 
  \begin{center}  
    \begin{tabular}{| c | c |}
    \hline
    $\quad p \quad $ & $ \quad \quad e_c \quad \quad $ \cr
    \hline
   5/1 &      0.211334\cr
   9/2 &      0.174269\cr
   4/1 &      0.135506\cr
   7/2 &      0.095959\cr
   3/1 &      0.057675\cr
   5/2 &      0.024877\cr
   2/1 &      0.004602\cr
   3/2 &      0.000026\cr
   1/1 &       $-$     \cr
    \hline
    \end{tabular}
  \end{center}
\end{table}

\section{Numerical simulations}

\llabel{NumSim}

We will now use the dynamical equations established in section~\ref{eqomt} to
simulate the final evolution of Mercury's spin by performing massive numerical
integrations.
The main goal is to illustrate the effects described in section~\ref{070803a},
in particular the probabilities of capture and escape from spin-orbit
resonances. 
Mercury geophysical models and parameters in use are those listed in
section~\ref{eqomt}. 
We recall here the most uncertain values: $ k_2 = 0.4 $, $ Q = 50 $ and $ \nu =
10^{-6} \mathrm{m}^2 \mathrm{s}^{-1} $.

\subsection{Simulations without planetary perturbations}

\llabel{070820z}

Before considering the effect of planetary perturbations we can 
test numerically the theoretical estimates of the capture probability given
by expressions (\ref{070803i}) and (\ref{070806a}).
Since capture in resonance is a statistical process we need to perform many
integrations with slightly different initial conditions.
For that purpose we ran 2000 simulations using a fixed eccentricity $ ( e =
0.206 ) $, initial rotation period of
\iniper~days, zero obliquity and different initial libration phase angles
with step-size of $ \pi / 1000 $~rad.
Results are listed in Table~\ref{TPcap}.
We can see that there is a good agreement between the theoretical previsions and
the numerical estimation of the probabilities. 
As discussed in sections~\ref{070803z} and~\ref{070803x} the probability of
capture when considering only the effect of tides is very small ($\sim 7 \% $
for the 3/2 resonance), while it becomes very important when core-mantle
friction is added ($100 \% $ for the 3/2 resonance).
They are also in conformity with those obtained in the previous studies by 
\citet{Goldreich_Peale_1967} and \citet{Peale_Boss_1977}. 
As they all noticed, when the effect from core-mantle friction is considered, the
probabilities of capture are greatly enhanced for all spin-orbit resonances.
In particular, capture in the 2/1 resonance also becomes 100\%, preventing a
subsequent evolution to the 3/2 resonance.

\begin{table}
  \caption{Capture probabilities in several spin-orbit resonances (in percentage). 
  In the left panel (T only) we consider the effect of tides alone, while in
  the right panel (T + CMF) both tides and core-mantle friction effects are
  taken into account. The first column ($ P_\mathrm{cap} $) refers to the
  theoretical estimation given by expression (\ref{070803g}), while the next
  column (num.) refers to the estimation obtained running a numerical simulation
  with 2000 close initial conditions, differing by $ \pi / 1000 $ in the
  libration angle. Planetary perturbations are not considered
  and we used a constant eccentricity $ e = 0.206 $. \llabel{TPcap}}    
  \begin{center}  
    \begin{tabular}{| c | c c | c c |} \hline
  & \multicolumn{2}{c|}{T only} &
    \multicolumn{2}{c|}{T + CMF}  \cr  
  $\; \quad p \quad \; $ & $ \; P_\mathrm{cap} \; $  &
   num.   & $ \quad P_\mathrm{cap} \; \; $  &  num.    \cr 
   & (\%)& (\%) & (\%) & (\%) \cr \hline
5/1& $-$ &  $-$ &   1.6 &   0.3   \cr 
9/2& $-$ &  $-$ &   3.1 &   1.3   \cr
4/1& 0.1 &  $-$ &   5.9 &   4.8   \cr
7/2& 0.1 &  0.1 &  11.4 &  10.9   \cr
3/1& 0.3 &  0.4 &  22.6 &  22.8   \cr  
5/2& 0.7 &  1.4 &  46.6 &  46.2   \cr
2/1& 1.8 &  1.7 & 100.0 & 100.0   \cr
3/2& 7.7 &  7.2 & 100.0 & 100.0   \cr \hline
    \end{tabular}
  \end{center}
\end{table}

\subsection{Inclusion of planetary perturbations}

\llabel{ecc.destab}

When planetary perturbations are taken into account, the eccentricity presents
chaotic variations with many excursions to higher and lower values than today
\citep[][Paper~I]{Laskar_1990,Laskar_1994,Laskar_2008}. 
It is then impossible to know its exact evolution at the time the planet first
encountered the spin-orbit resonances. 
A statistical study with many different orbital solutions is the only possibility to get a
global picture of the past evolution of the spin of Mercury.
In Paper~I we performed such a study by integrating 1000 orbits over 4~Gyr in the
past starting with very close initial conditions.
This statistical study was only made possible by the use of the averaged
equations for the motion of the Solar System \citep{Laskar_1990,Laskar_1994}.

In figure~\ref{ecc5} we show five examples of the eccentricity evolution through
the 4.0\,Gyr.
We choose some cases illustrative of the chaotic behavior, where we can see that
the eccentricity can be as small as zero, but it can also reach values as high
as 0.5.
In some cases the eccentricity can remain within $[0.1,0.3]$ throughout the
evolution, while in other cases it can span the whole interval $[0,0.5]$
\citep{Laskar_2008}.

Owing to the chaotic evolution, the density function of the 1000 solutions over
4\,Gyr is a smooth function (Fig.\ref{LA04}), well approximated by a Rice probability
distribution \citep{Laskar_2008}.
The eccentricity excursions to higher values allow the planet to cross the 3/2
resonance several times, and thus increase the probability of capture.
This behavior becomes very important if the evolution is driven by tidal
friction alone.
Even though the probability of capture in a single crossing of the 3/2
spin-orbit resonance is only around 7\%, multiple crossings increase it up
to 55\% (Paper~I).

\subsection{Planetary perturbations with core-mantle friction}

In presence of an efficient core-mantle friction the multiple crossings of the
3/2 resonance are no longer needed, since the capture in this resonance after a
single crossing is already 100\% (Fig.\ref{probe}).
Nevertheless, eccentricity excursions to lower values can destabilize the
equilibrium in any spin-orbit resonance different from the 1/1
(Tab.\ref{Tabb}).
This effect was already present when the core-mantle friction was not considered
(Paper~I), but with small influence on the results, while here it
becomes of capital importance.  
Indeed, it may allow the evasion from previous captures in higher order
resonances than the 3/2 and permit subsequent evolution to the present observed
spin state.

In order to check this new scenario, we have performed a statistical study of
the past evolutions of Mercury's orbit, with the integration of the same 1000
orbits over 4\,Gyr in the past used in Paper~I.
We now additionally include the effect of core-mantle friction as described by
\citet{Goldreich_Peale_1967}, i.e., we will consider the 
full dynamics of the spin governed by Eq.(\ref{070803b}).

Assuming an initial rotation period of Mercury of 10\,h, we estimated that the
time needed to de-spin the planet to the slow rotations would be about 300
million years. 
We will then start our integrations already in the slow-rotation regime, with a
rotation period of 10~days ($ \omega \simeq 8.8 n $), zero obliquity and
a starting time of $-4$\,Gyr, although these values are not critical.  
In Table~\ref{TPcap2} we show the amount of captures for each resonance
at the end of the simulations (column ``final'').
We also list the resonances in which the spin was first captured before being
destabilized (column ``1$^{st}$cap.'') and we recall the results obtained for a
constant eccentricity $ ( e = 0.206 ) $ and in Paper~I, with
a model without core-mantle friction.

\begin{table}
  \caption{Capture probabilities in several spin-orbit resonances (in percentage). 
  We performed a statistical study of the past evolutions of Mercury's spin,
  with the integration of 1000 orbits over 4\,Gyr, a initial rotation period of
  10\,days and zero obliquity.
  In the ``1$^{st}$cap.'' column we list the
  resonances in which the spin was first captured (before being destabilized).
  In the ``final'' column we list the results after the full 4\,Gyr of
  simulations. For comparison we also list the results obtained with a constant
  eccentricity $ e = 0.206 $ (``const.'') and the final results obtained in
  Paper~I (``C\&L04'' ), with a model without core-mantle
  friction. \llabel{TPcap2}} 
  \begin{center}  
    \begin{tabular}{| c | c c c c |} \hline
    & \multicolumn{4}{c|}{number of captures} \cr
$ \; \quad p \quad \; $ & $\;$ const. $\;$ 
& $ \; 1^{st}$cap. $\;$ & $\;$ final $\;$ & $\;$ C\&L04 $\;$   \cr 
    &   (\%)  &   (\%)  &   (\%) &   (\%) \cr \hline
 6/1&  $-$ &  0.1 &  $-$ &  $-$ \cr
11/2&  $-$ &  0.4 &  $-$ &  $-$ \cr
 5/1&  0.3 &  1.3 &  $-$ &  $-$ \cr
 9/2&  1.3 &  2.7 &  $-$ &  $-$ \cr
 4/1&  4.7 &  5.3 &  $-$ &  $-$ \cr
 7/2& 10.3 &  8.7 &  4.7 &  $-$ \cr
 3/1& 19.0 & 15.5 & 11.6 &  $-$ \cr  
 5/2& 29.8 & 26.5 & 22.1 &  $-$ \cr
 2/1& 34.6 & 31.2 & 31.6 &  3.6 \cr
 3/2&  $-$ &  8.1 & 25.9 & 55.4 \cr 
 1/1&  $-$ &  0.2 &  3.9 &  2.2 \cr 
none&  $-$ &  $-$ &  0.2 & 38.3 \cr \hline
\end{tabular}
  \end{center}
\end{table}

After running 1000 trajectories we observe that the spin of Mercury preferably
chooses one of the three final configurations: 5/2, 2/1 or 3/2 (Tab.\ref{TPcap2}).
With 26\% of captures, the present configuration no longer represents the most probable final
outcome, as it was in absence of core-mantle friction (Paper~I).
However, it is still among the most probable scenarios, the alternatives
receiving comparable amounts of captures (22\% and 32\% respectively for the 5/2
and the 2/1 resonances).
The 5/2 and the 2/1 spin-orbit resonances benefit from the fact that the planet
must cross them first. 
On the other hand, the 3/2 resonance is more
stable and the chances of capture are higher when crossed.


Since the eccentricity of Mercury 4.0\,Gyr ago can be around 0.4
(e.g. Fig.\ref{ecc5}), at the moment of the first encounter with the
spin-orbit resonances, capture in
resonances as high as the 6/1 can occur (Fig.\ref{probe}).
Because the probability of capture is small and because there are not
many orbital solutions reaching such high values for the eccentricity, we only
count about 10\% of captures in resonances above or equal to the 4/1 (Tab.\ref{TPcap2}).
Once captured, these equilibria can be maintained as long as the eccentricity
remains above the respective critical values (Tab.\ref{Tabb}).

Contrary to the results predicted for a constant eccentricity
(Tab.\ref{TPcap2}), we also registered a few trajectories directly
captured in the 3/2 resonance just after the first passage through the
resonance area.
When we used the present value of the eccentricity ($ e = 0.206 $), the 3/2
resonance could not be attained because for that value
capture probability in the 2/1 resonance is 100\% (Fig.\ref{probe}).
However, for eccentricity values lower than about 0.19, the probability of
capture in this resonance decreases, as well as for higher order resonances.
For instance, when the eccentricity is 0.09, capture in the 2/1 resonance drops
to 50\%.
Thus, since the eccentricity of Mercury is varying, it may happen that about
4.0\,Gyr ago its value was much lower than today and the spin managed to avoid
all the spin-orbit resonances higher than the 3/2 and was directly captured in
the present observed configuration.
We estimate nevertheless that the probability for this scenario to occur is
very low, only about 8\% (Tab.\ref{TPcap2}).

\subsection{Critical eccentricities}

Over 4.0\,Gyr of evolution the eccentricity has many chances of
experiencing a period of very small values (e.g. Fig.\ref{ecc5}).
Even when a period of low eccentricity does not last for a long time, a single
passage of the eccentricity below a critical value (Tab.\ref{Tabb}) can be enough
to destabilize the corresponding spin-orbit resonance.

All orbital solutions were generated starting from initial conditions close to
the present values \citep[see][]{Laskar_2008}, and therefore converge to the
same final evolution. 
The eccentricity behavior is thus identical for the last 50-60\,Myr
(Fig.\ref{ecmerc}), before which the chaotic diffusion dominates
(Fig.\ref{econv}). 
During the last 50\,Myr the eccentricity certainly reached values lower than
0.13, thus the 4/1 and above spin-orbit resonances cannot
represent a possible final outcome for Mercury ($ e_{4/1} \approx 0.136 $). 
For the 7/2 and lower order spin-orbit resonances, capture until the present
day is not forbidden by the last 50\,Myr of Mercury's evolution, but depends on
the true orbital evolution of the eccentricity (Fig.\ref{econv}).
The higher is the critical eccentricity, the lower is the probability of
remaining trapped, because more orbital solutions will come below this value.

In figure~\ref{eccdist} we plot the cumulative distribution of the minimal eccentricities
attained for each one of the 1000 orbital solutions that we used.
We also mark with straight lines the critical values of the eccentricity for each
spin-orbit resonance (Tab.\ref{Tabb}) and use dots to represent the amount of
captures obtained numerically for spin-orbit resonances that are still stable
below each critical value of the eccentricity.
Since a large amount of the orbital solutions experience at least one episode with an
eccentricity below 0.05 ($ \log_{10} e \approx -1.3 $), about 84\% of the final
evolutions will end in the 5/2 spin-orbit resonance or lower (Tab.\ref{TPcap2}).
By comparing the eccentricity instability thresholds for each spin-orbit
resonance with the amount of captures obtained numerically below that resonance
we see that there is a good agreement, suggesting a strong
correlation between the orbital evolution of the eccentricity and the percentage
of captures in each resonance. 
The reason why there is not full agreement between the two is because
spin-orbit resonances below critical values of the eccentricity can also be
attained by trajectories that escaped capture in higher order resonances, that
is, they can be attained even if the eccentricity is never below the critical
value for that resonance (Fig.\ref{probe}).

When comparing the results after 4.0\,Gyr with those after the first capture, we
verify that the 5/2 resonance (and above) lose a significant amount of previously
captured solutions.
The amount of orbits captured in the 2/1 resonance remains roughly the same,
because the number of trajectories quitting this resonance is more or less
compensated by the incoming trajectories from higher order resonances.
The 3/2 is the real winner of this transition process, as the amount of
trajectories that end in this last configuration is about 4 times larger
than it was initially.
An identical scenario was already observed in Paper~I,
except that only a few captures occurred in spin-orbit resonances higher than the
2/1 and they were all subsequently destabilized (Tab.\ref{TPcap2}).

As in Paper~I, we also noticed about 4\% of the trajectories
captured in the 1/1 spin-orbit resonance (Tab.\ref{TPcap2}).
Since the probability of capture in the 3/2 spin-orbit resonance is almost 100\%
even for very low values of the eccentricity (Fig.\ref{probe}), the major
possibility of evolving into the 1/1 resonance is by destabilizing the 3/2.
This becomes a possibility if the eccentricity is almost zero, that is, for $ e
< 3 \times 10^{-5} $ (Tab.\ref{Tabb}).

\subsection{Different scenarios of evolution}

\llabel{diffscenarios}

The critical eccentricity needed to destabilize the 2/1 spin-orbit resonance is
$ e_{2/1} \approx 0.0046 $ (Tab.\ref{Tabb}).
Whenever the orbital eccentricity is below this value, the spin will then
evolve towards the 3/2 resonance or below (Fig.\ref{wnecc}).
However, this is not the only possibility of achieving this last configuration
if the planet was first captured in a higher-order resonance.
Indeed, as discussed for the $1^{st}$ capture column (Tab.\ref{TPcap2}), if the
eccentricity is lower than 0.19 at the time the planet crosses the 2/1 resonance
(Fig.\ref{probe}), the chances of capture are lower than 100\%, opening some
space for subsequent evolution to the 3/2 resonance.
For instance, for a previous capture in the 5/2 resonance, an eccentricity of
$ e_{5/2} \approx 0.025 $ will destabilize it and produce a capture probability
of only 14.4\% in the 2/1 resonance, i.e., when the 5/2 resonance is
destabilized there is about 85\% of chance of ending in the 3/2 present
configuration (Fig.\ref{wnecc}).

In order to exemplify the multitude of possible evolutionary scenarios, we
performed another kind of experiment.
Adopting a particular orbital solution, which presents a gradual
decrease in the eccentricity (Fig.\ref{ecc5}a), we integrated close initial
conditions for the spin.
Since for this orbital solution the eccentricity is high at the time of the
first encounter, there is a great chance of capturing the spin in a spin-orbit
resonance with $ p > 5/2 $.
In figure~\ref{wnecc} we plot the behavior of four trajectories, each one
initially captured in a different spin-orbit resonance, when the eccentricity
approaches a zone of very low values.
As expected, the spin-orbit resonances are sequentially abandoned as the
eccentricity assumes small values.
In particular, we observe that the resonances are quit immediately after the
eccentricity is below the critical values listed in Table~\ref{Tabb}.
After being destabilized, the spin can evolve directly to the present 3/2
configuration, or can be trapped in an intermediate spin-orbit resonance.
In this example, the eccentricity become lower than $ e_{2/1} \approx 0.0046 $
around $-1.79$\,Gyr and captures in the 2/1 resonance become destabilized after
that date. 

We purposely plot one situation, where the spin does not
end in the 3/2 resonance, however.
In this case, at the moment the eccentricity becomes lower than $ e_{2/1} $, the
spin is still captured in the 5/2 resonance.
This resonance logically becomes destabilized and the rotation rate decreases.
Nevertheless, at the moment the spin encounters the 2/1 resonance the
eccentricity is again higher than $ e_{2/1} $, and therefore there is a chance of
capture in this resonance, preventing a subsequent evolution toward the 3/2
state (Fig.\ref{wnecc}).


\subsection{Constraints on the orbital evolution}

We have seen in previous sections that there is an important correlation between
the minimal eccentricity attained by Mercury through its orbital evolution and
the probability of capture in a given resonance (Fig.\ref{eccdist}).
The lower is the minimal eccentricity, the higher is the probability of achieving
a low order spin-orbit resonance.
In particular, each time the eccentricity descends below a given
critical value for a spin-orbit resonance (Tab.\ref{Tabb}), the spin will evolve
into a lower resonance.

Since we know the distribution of the minimal eccentricities (Fig.\ref{eccdist}), we
can estimate the probability of ending in a specific spin-orbit resonance
given the value of the minimal eccentricity of a considered orbit, $ P_{\mathrm{cap}/e_c} $.
For that purpose we eliminate all the trajectories for which the minimal eccentricity is
above the critical value (Tab.\ref{Tabb}) and then count the
number of captures in each resonance for the remaining orbital solutions.
Results are listed in Table~\ref{TabC1}.
While for an arbitrary orbital solution the probability of capture in the present 3/2
spin-orbit resonance is only 25.9\%, this value rises to 55.1\% if we assume
that the eccentricity of Mercury was below $ e_{5/2} \approx 0.025 $ at some time in the
past, or even up to 73.2\% if the eccentricity descends below $ e_{2/1} \approx
0.0046 $.
Results for the critical eccentricity $ e_{4/1} \approx 0.136 $ are the same as
the global results shown in Table~\ref{TPcap2}, because the eccentricity of
Mercury was below that value in the most recent 50~Myr, where the chaotic
behavior is not significant (Fig.\ref{econv}).
Notice also that there are always a few captures left in
resonances above the corresponding critical value.
This can be explained by the same effect described in the last paragraph of
section~\ref{diffscenarios} and illustrated in figure~\ref{wnecc}.

\begin{table}
  \caption{Capture probabilities in spin-orbit resonances (in percentage), when the
  eccentricity descends below a given critical value (Tab.\ref{Tabb}).
   \llabel{TabC1}}
  \begin{center}  
    \begin{tabular}{| c | c c c c c c |} \hline
$ \; \quad p \quad \; $ & $ \; e_{3/2} \; $ & $ \; e_{2/1} \; $ & $ \; e_{5/2} \; $ 
& $ \; e_{3/1} \; $ & $ \; e_{7/2} \; $ & $ \; e_{4/1} \; $  \cr \hline
 7/2&   $-$ &  $-$ &  1.2 &  2.6 &  3.8	&  4.7 \cr
 3/1&   $-$ &  2.0 &  4.2 &  7.2 & 10.7	& 11.6 \cr  
 5/2&   $-$ &  3.2 &  5.2 & 16.0 & 20.7	& 22.1 \cr
 2/1&   3.1 &  5.4 & 23.9 & 30.5 & 32.6	& 31.6 \cr
 3/2&  25.0 & 73.2 & 55.1 & 37.2 & 27.7	& 25.9 \cr 
 1/1&  68.8 & 15.4 &  9.7 &  6.1 &  4.2	&  3.9 \cr 
none&   3.1 &  0.7 &  0.5 &  0.3 &  0.2	&  0.2 \cr \hline
\end{tabular}
  \end{center}
\end{table}

Inversely, since we know that the rotation of Mercury is presently captured in
the 3/2 spin-orbit resonance, we can estimate the probability for the
eccentricity to have descended during its past evolution below a specific
critical level, $ P_{e_c} $. 
Using conditional probabilities, we have then
\be
P_{e_c} = \frac{P_{\mathrm{cap}/e_c} \times P_\mathrm{orb}}{P_\mathrm{cap}} \ ,
\llabel{080620a}
\ee
where $ P_{\mathrm{cap}/e_c} $ is the probability of ending in a specific spin-orbit
resonance given the value of the minimal eccentricity (Tab.\ref{TabC1}), $
P_\mathrm{orb} $ the probability for the eccentricity to reach that minimal
eccentricity (Fig.\ref{eccdist}) and $ P_\mathrm{cap} $ the global capture probability
in the specific spin-orbit resonance (Tab.\ref{TPcap2}).
Results for the 3/2 spin-orbit resonance are given in Table~\ref{TabC2}.
These probabilities for the orbital evolution of the eccentricity are the same
as if we select only the evolutions that finished in the 3/2 spin-orbit
resonance and then look at the minimal eccentricity distribution.
From the above analysis we conclude that there is a strong probability that 
the eccentricity of Mercury reached very low values; in particular there is
about a 77\% chance that it descended below $ e_{2/1} \approx 0.0046 $ (but
only a 3\% chance of going below $ e_{3/2} \approx 0.00003 $).

\begin{table}
  \caption{Probability for the eccentricity of Mercury to have descended below a
  specific critical level ($P_{e_c}$) given that its rotation is captured in the
  3/2 spin-orbit resonance today. Results for each critical eccentricity $ e_c $
  (Tab.\ref{Tabb}) are obtained from Eq.(\ref{080620a}) with $P_\mathrm{cap} =
  25.9$\% (Tab.\ref{TPcap2}).
   \llabel{TabC2}}
  \begin{center}  
    \begin{tabular}{| c | c c c c c c |} \hline
$ \; \quad P \quad \; $ & $ \; e_{3/2} \; $ & $ \; e_{2/1} \; $ & $ \; e_{5/2} \; $ 
& $ \; e_{3/1} \; $ & $ \; e_{7/2} \; $ & $ \; e_{4/1} \; $  \cr \hline
$P_{\mathrm{cap}/e_c}$ &25.0 & 73.2 & 55.1 & 37.2 & 27.7 &  25.9 \cr 
$P_\mathrm{orb}      $ & 3.2 & 27.1 & 40.3 & 64.5 & 99.7 & 100.0 \cr \hline
$P_{e_c}             $ & 3.1 & 76.6 & 85.7 & 92.6 & 98.8 & 100.0 \cr \hline
\end{tabular}
  \end{center}
\end{table}

\section{Conclusions}

Due to the increasing evidence of a molten core inside Mercury
\citep{Ness_etal_1974,Margot_etal_2007}, viscous friction at the core-mantle
boundary is expected and its consequences to the spin must be taken into account.
An important consequence is a considerable increase in the probability of
capture for all spin-orbit resonances; in particular, for the 2/1 and the 3/2 it
can reach 100\% \citep{Peale_Boss_1977}.
Since it is believed that Mercury's initial rotation was much faster than today,
a destabilization mechanism is then required to allow the planet to escape from the
2/1 and higher order resonances and subsequently evolve to the present observed
3/2 configuration. 

With the consideration of the chaotic evolution of the eccentricity of
Mercury we show that such destabilization mechanism exists whenever the
eccentricity becomes smaller than a critical value for each spin-orbit resonance
(Tab.\ref{Tabb}). 
This mechanism was already described in Paper~I,
but becomes of capital importance when core-mantle friction is taken into
account.
There are two main possibilities to evolve into the 3/2 configuration:

\vskip.2cm
\begin{itemize}
\item The eccentricity becomes lower than the critical value for the 2/1
spin-orbit resonance ($ e_{2/1} \approx 0.005 $) and evolves into the 3/2.

\item The eccentricity becomes lower than the critical value for a higher order
resonance than the 2/1, and then crosses this resonance with an eccentricity
lower than $ e < 0.19 $. This allows a non zero probability of escaping the 2/1
resonance, and subsequent evolution into the 3/2. 
\end{itemize}

\vskip.2cm
The other mechanism of capturing in the 3/2 resonance described in
Paper~I, consisted in a returning to the 3/2 spin-orbit
resonance after an increase in the eccentricity.
This effect is not as important when we take into account core-mantle
friction, since the most part of the trajectories are captured in resonance
after a single passage.

After running 1000 orbital solutions, starting from 4\,Gyr in the past until they reached
the present date, the spin of the planet was captured in a spin-orbit resonance
99.8\% of the time. 
The main resonances to be filled and the respective probability  were
(Tab.\ref{TPcap2}): 
\be
P_{5/2}=22.1\%,\quad  P_{2/1}=31.6\%,\quad P_{3/2}=25.9\% \ .
\ee
Although in this case the present configuration no longer represents the most
probable final outcome, as it was in absence of core-mantle friction (Paper~I), 
it is still among the most probable scenarios.

Moreover, if we assume  that at some time in the past, the eccentricity of Mercury 
becomes lower than  $ e_{5/2} \approx 0.025 $ or $ e_{2/1} \approx 0.005 $
respectively,
the probability of reaching the 3/2 spin-orbit resonance rises to
55\% and 73\% respectively (Table \ref{TabC1}).
Given that Mercury is presently trapped in the 3/2 configuration, we can also 
estimate that the eccentricity of Mercury has known at least one period of very
low eccentricity during its past evolution, with about 86\% and 77\% of chances
of being below $ e_{5/2} \approx 0.025 $ and $ e_{2/1} \approx 0.005 $
respectively (Table \ref{TabC2}).

The probability of capture in the 3/2 resonance can also be increased if the
orbital eccentricity experiences more periods near zero.
This can be achieved if we use direct integration of the Solar System instead
of the averaged equations, because the true eccentricity is expected to undergo
some additional small variations around the value obtained for the averaged equations
\citep{Laskar_2008}.
Alternatively, the probability of capture in the 3/2 resonance can still be
increased if we are able to increase the critical eccentricities that
destabilize spin-orbit resonances (Tab.\ref{Tabb}). 
This can be achieved if the tidal dissipation is stronger ($ k_2 > 0.4 $ and/or
$ Q < 50 $) or if $ C_{22} < 1.0 \times 10^{-5} $ (Eq.\ref{080319a}).

Lower values for the core effective viscosity, $ \nu $, will not change the
critical eccentricities, but will decrease the amount of captures for all
spin-orbit resonances.
As a consequence, it becomes easier to escape from the capture in spin-orbit resonances,
and all those trajectories that also escape the 3/2 resonance can be later
trapped there when the eccentricity experiences a period with $ e > 0.325 $
(Paper~I).
We then believe that the true scenario for the evolution of the spin of Mercury
may be somewhere between the scenario described here, with an efficient core-mantle
friction effect, and the scenario described in Paper~I, for a
total absence of core-mantle friction.
In the future, different dissipative parameters and models could be tested as
well as the effect of the obliquity, that was supposed to be zero in the present
study.

\section*{Acknowledgments}

The authors thank S.J. Peale for discussions.
This work was supported by the Funda\c{c}\~{a}o Calouste Gulbenkian (Portugal),
Funda\c{c}\~{a}o para a Ci\^{e}ncia e a Tecnologia (Portugal), and by PNP-CNRS
(France).

\bibliographystyle{elsarticle-harv}      
\bibliography{correia}   


\begin{figure}
  \begin{center}
    \includegraphics*[width=9cm]{\figpath 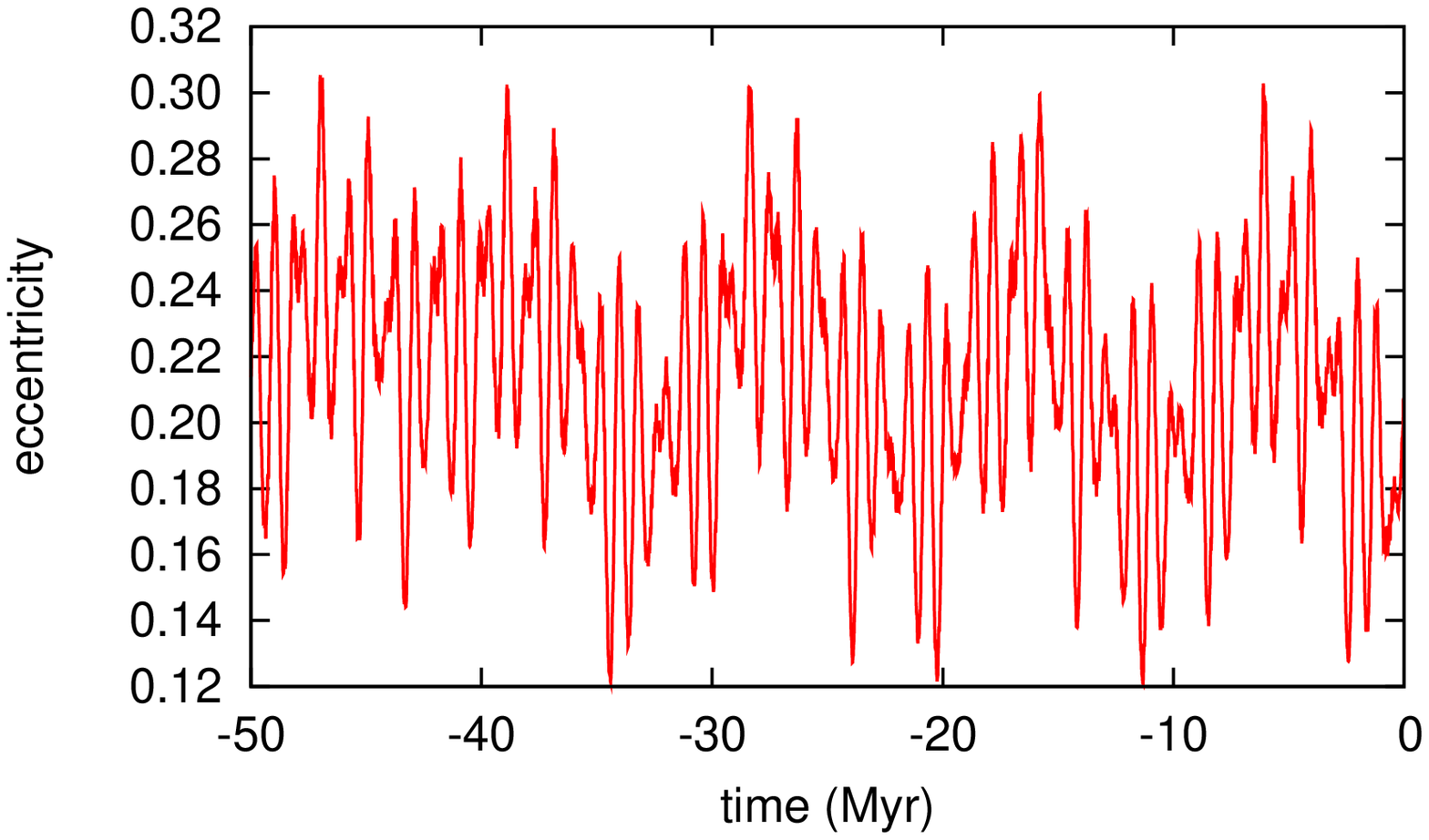}
  \caption{Evolution of Mercury's eccentricity over 50 Myr in the past 
  \citep{Laskar_etal_2004M,Laskar_etal_2004E}. \llabel{ecmerc} }
  \end{center}
\end{figure}

\begin{figure*}
  \begin{center}
    \includegraphics*[\wh=16cm]{\figpath 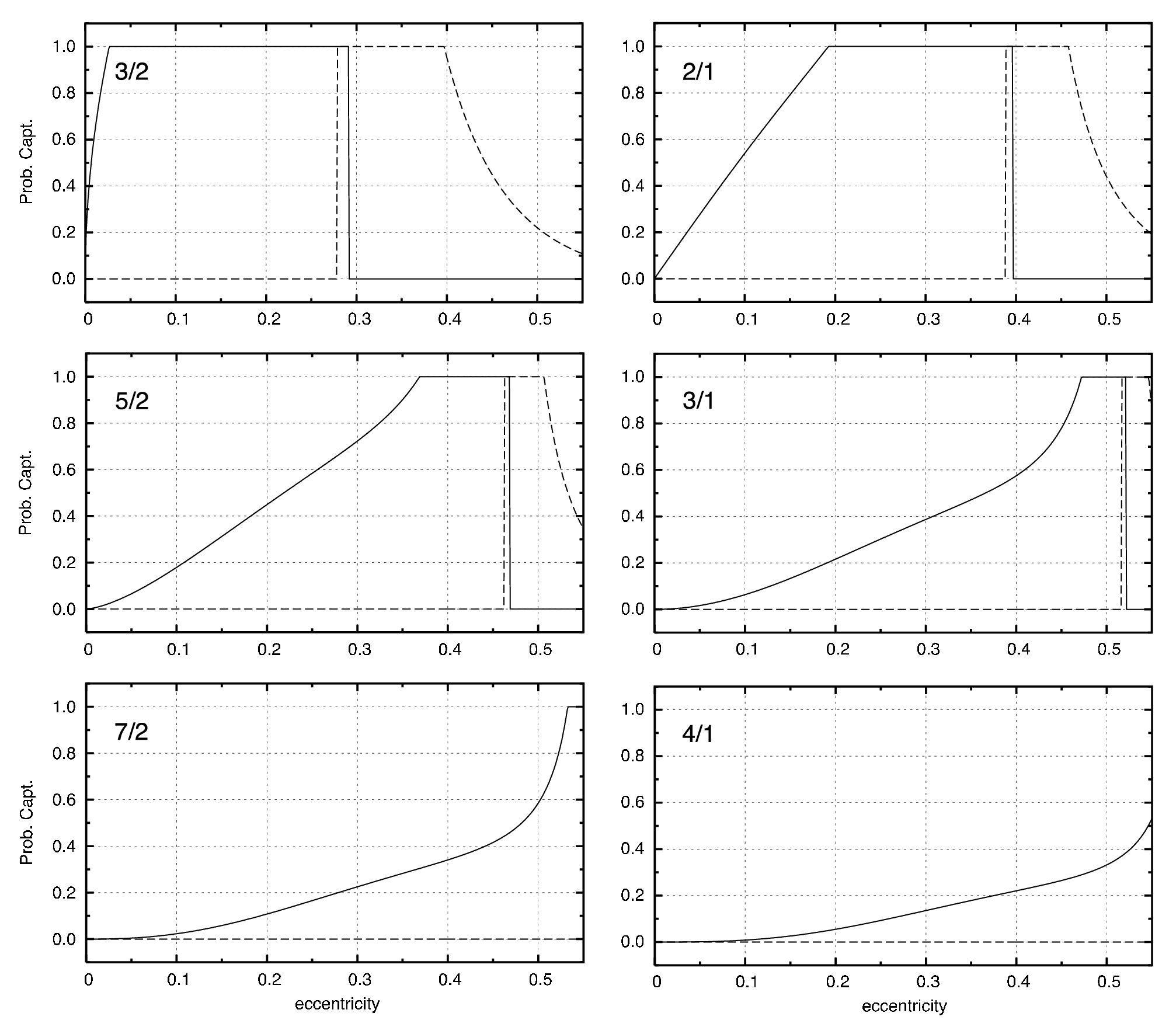}
  \caption{Probability of capture in some spin-orbit resonances for different
  values of the eccentricity, under the effect of tides and core-mantle friction
  (Eq.\ref{070806a}).   
  The dashed line corresponds to a planet increasing its spin 
  from slower rotation rates, while the solid line corresponds to a planet
  de-spinning from faster rotation rates. 
  For all resonances, capture probability is 100\% whenever the eccentricity is
  close to the equilibrium value for the rotation rate, $ \omega / n = E(e) $
  (Eq.\ref{070803h}). 
  It suddenly decays to zero when the equilibrium rotation rate falls outside
  the resonance width, i.e., the tidal evolution prevents the planet from
  crossing the resonance. \llabel{probe} }
  \end{center}
\end{figure*}

\begin{figure}
  \begin{center}
    \includegraphics*[width=9cm]{\figpath 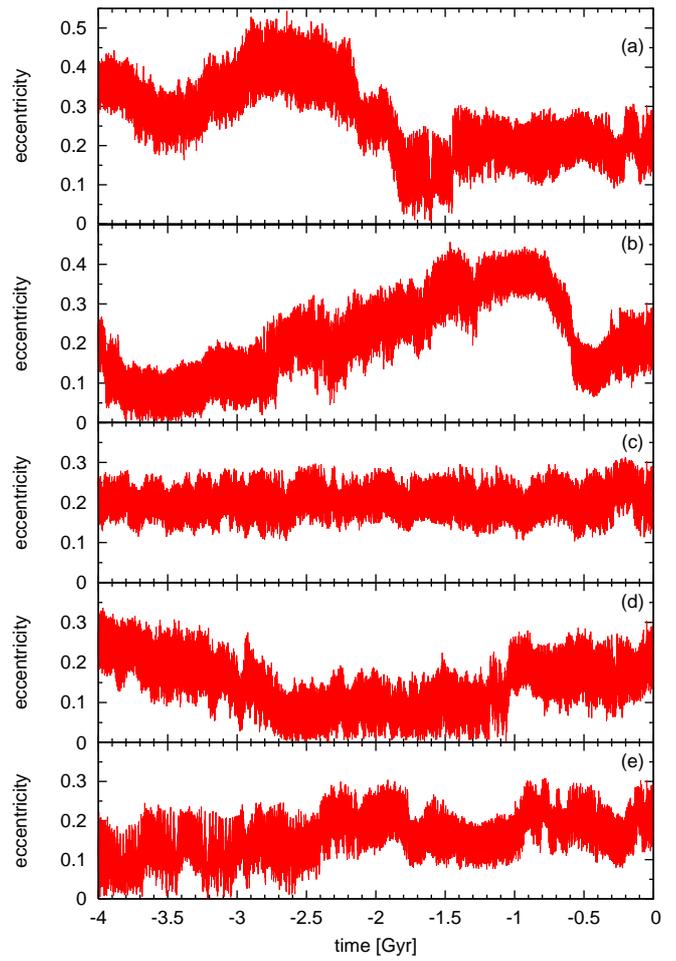}
  \caption{Some examples of the possible variations of the
  eccentricity of Mercury through the past 4.0\,Gyr.
  The eccentricity can be as small as zero, but it can also reach values as high
  as 0.5. In some cases the eccentricity can remain within $[0.1,0.3]$
  throughout the evolution, while in other cases it can span the whole interval
  $[0,0.5]$.
  All these solutions converge to the known recent evolution of the planet's
  orbit (Fig.\ref{econv}). 
  \llabel{ecc5} }
  \end{center}
\end{figure}

\begin{figure}
  \begin{center}
    \includegraphics*[\wh=8.5cm]{\figpath 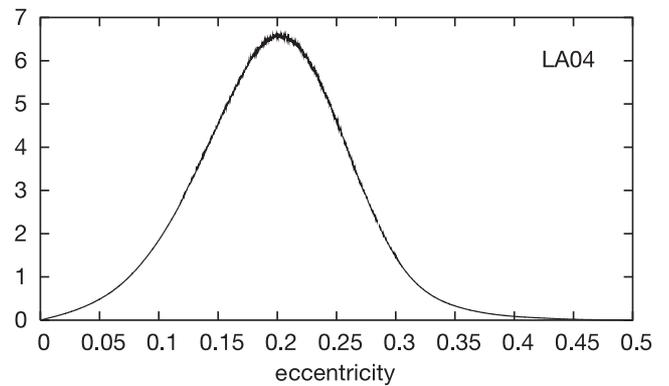}
  \caption{Probability density function of Mercury's eccentricity
  (Paper~I). Values are
  computed over 4\,Gyr for the numerical integration of the secular
  equations \citep{Laskar_etal_2004M,Laskar_etal_2004E,Laskar_2008} for 1000 close initial
  conditions (LA04). The mean value of the eccentricity is $ \bar e = 0.198 $.
  \llabel{LA04}}  
  \end{center}
\end{figure}

\begin{figure}
  \begin{center}
    \includegraphics*[height=8.5cm,angle=\vangle]{\figpath 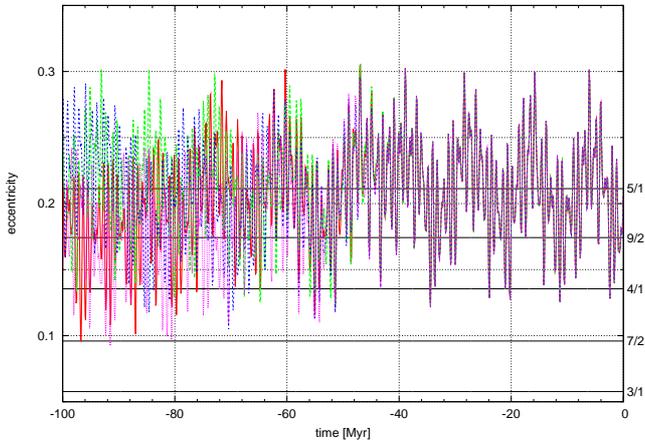}
  \caption{Some examples of the recent evolution of the
  eccentricity of Mercury. During the last 50-60\,Myr all the orbits present the
  same evolution, before which the chaos effect takes place.  
  Horizontal lines correspond to the critical eccentricities that
  destabilize the equilibrium in a given spin-orbit resonance (Tab.\ref{Tabb}).
  During the last 50\,Myr the eccentricity was certainly below 0.13, thus the
  4/1 resonance is not a possible final outcome for Mercury. For some
  orbits the eccentricity is below 0.09 in the last 100\,Myr and the 7/2
  resonance was also destabilized. However, since this scenario is not true for
  all orbital solutions, we may expect a few final evolutions captured in this
  last configuration (Tab.\ref{TPcap2}).
  \llabel{econv} }
  \end{center}
\end{figure}

\begin{figure}
  \begin{center}
    \includegraphics*[width=8.5cm]{\figpath 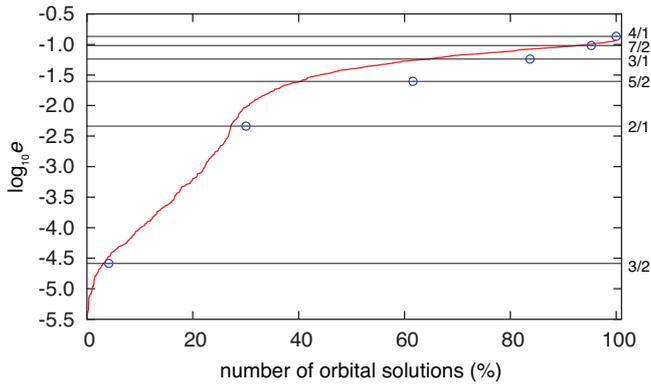}
  \caption{Cumulative distribution of the minimal eccentricities attained for the
  1000 orbital solutions that we used. Straight lines represent the
  critical values of the eccentricity for each spin-orbit resonance
  (Tab.\ref{Tabb}), while dots represent the amount of captures obtained
  numerically for spin-orbit resonances that are still stable below each
  critical value of the eccentricity (Tab.\ref{TPcap2}).
  \llabel{eccdist} }
  \end{center}
\end{figure}

\begin{figure*}
  \begin{center}
    \includegraphics*[height=17.5cm,angle=\vangle]{\figpath 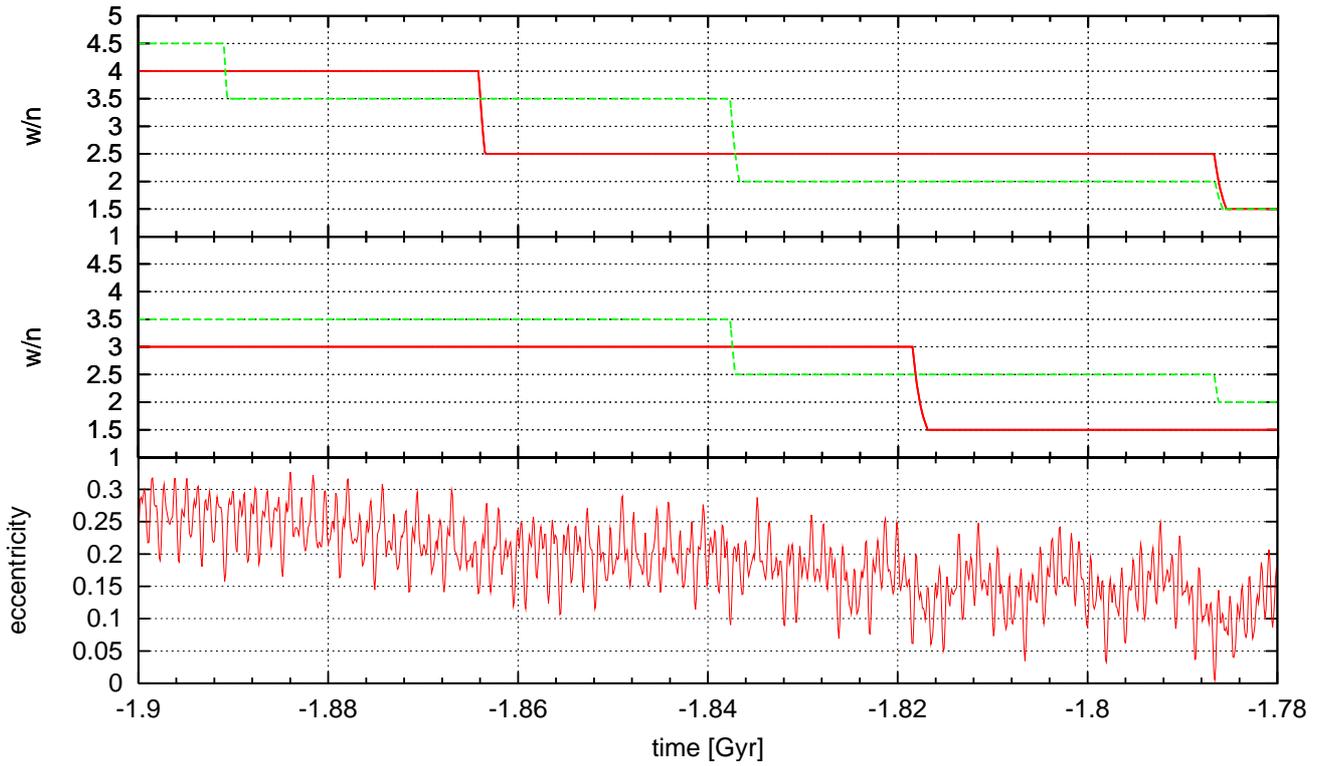}
  \caption{Four possible final evolutions for the spin of Mercury. Adopting a
  particular orbital solution, which presents a gradual decrease in the
  eccentricity (Fig.\ref{ecc5}a), we integrate close initial conditions for the
  spin. We observe that spin-orbit resonances are quit immediately after the
  eccentricity is below the critical values listed in Table~\ref{Tabb}. After
  being destabilized, the spin can evolve directly to the present 3/2
  configuration, or can be trapped in an intermediate spin-orbit resonance. We
  purposely left one situation where the spin does not end in the 3/2
  resonance. At the moment the eccentricity becomes lower than $ e_{2/1} \approx
  0.0046 $, the spin is still captured in the 5/2 resonance. This resonance is
  then destabilized, but when the spin encounters the 2/1 resonance the
  eccentricity is already higher than $ e_{2/1} $. Thus, there is a chance of
  capture in this resonance, preventing a subsequent evolution toward the 3/2
  state.
 \llabel{wnecc} }
  \end{center}
\end{figure*}

\end{document}